\documentclass[12pt]{iopart}
\usepackage{amsfonts}
\usepackage{amssymb}
\usepackage{graphicx}
\usepackage{epsfig}
\usepackage{color}

\begin{document}

\title{On exact time-averages of a massive Poisson particle}
\author{
\mbox{W A M Morgado$^1$},
\mbox{S M Duarte Queir\'{o}s$^{2,3}$}
and \mbox{D O Soares-Pinto$^4$}}

\address{$^1$ Department of Physics, PUC-Rio and \\
National Institute of Science and Technology for Complex Systems \\
Rua Marqu\^es de S\~ao Vicente 225, G\'avea, CEP 22453-900 RJ, Rio de
Janeiro, Brazil}
\address{$^2$ Centro de F\'{\i}sica do Porto \\
Rua do Campo Alegre, 687, 4169-007 Porto, Portugal}
\address{$^3$ Istituto dei Sistemi Complessi, CNR \\
Via dei Taurini 19, 00185 Roma, Italy}
\address{$^4$ Instituto de F\'{\i}sica de S\~{a}o Carlos \\ Universidade de S\~{a}o Paulo,
PO Box 369, CEP 13560-970 SP, S\~{a}o Carlos, Brazil.}

\eads{\mailto{welles@fis.puc-rio.br},\mailto{sdqueiro@gmail.com},%
\mailto{diogo.osp@ursa.ifsc.usp.br}}

\begin{abstract}
In this work we study, under the Stratonovich definition, the problem of the damped oscillatory massive particle subject to a heterogeneous Poisson noise characterised by a rate of events, $\lambda (t)$, and a magnitude, $\Phi $, following an exponential distribution. We tackle the problem by performing exact time-averages over the noise in a similar way to previous works analysing the problem of the Brownian particle. From this procedure we obtain the long-term equilibrium distributions of position and velocity as well as analytical asymptotic expressions for the injection and dissipation of energy terms. Considerations on the emergence of stochastic resonance in this type of system are also set forth.

\end{abstract}


\vspace{2pc} \noindent\textit{Keywords}: Stochastic processes, Rigorous
results in statistical mechanics
\newline
\submitto{J. Stat. Mech. : Theor. Exp.}

\maketitle

\section{Introduction}

Stochastic processes have long surpassed the limits of a mere probability
theory subject establishing itself as a topic of capital importance in
various disciplines which go from molecular motors to internet traffic
analysis~\cite{livro_vankampen,Hanggi2009,Leland1994}. In what concerns
statistical mechanics, we owe its introduction to the study of Brownian
motion by means of the Langevin equation~\cite{VanKampen2005}. Explicitly, a
random term was added to the classical laws of movement in order to emulate
the collisions between the particle under study and the particles of the
supporting medium assuming that the jolts between particles cannot be
deterministically written down~\cite{livro_nelson,VanKampenOppenheim_86}.
This approach was promptly adopted in other problems, particularly those
coping with systems that exhibit a large number of degrees of freedom.
Therehence, the allocation to the noise of the microscopic features and
playing interactions of the system has become a common practice. Due to the
accurate descriptions of diverse systems, the use of the Wiener process
(driftless Brownian motion)~\cite{livro_ioannis,livro_revuzyor} has assumed
a leading role in widespread fields. However, this corresponds to a very
specific sub-class in the wider L\'{e}vy class of stochastic process~\cite%
{Bouchaud1990}. Specifically, the Wiener process corresponds to the case of
a stochastic process for which the increments are independent, stationary
(in the sense that the distribution of any increment depends only on the
time interval between events) associated with a Gaussian distribution thus
having all statistical moments finite. Moreover, its intimate relation to
the Fokker-Planck Equation has put the Wiener process and its adaptive
processes in the limelight.

Despite the ubiquity of the Wiener process, several other processes, either
in Nature or man made, are quite distinct~\cite%
{Baule2009,Anteneodo2010a,Dykman2010}. Particularly, they can be associated
with (compound) Poisson processes in which the independent increments befall
upon a rate $\lambda \left( t\right)$ with its magnitude related to a
certain probability density function. For this kind of noise, it was
previously shown that the traditional Fokker-Planck approach cannot be
applied because the Kramers-Moyal moments from the third order on do not
vanish~\cite{Hanggi1981}. Consequently, the equation for the evolution of
the probability density function cannot be exactly written as a second-order
differential equation, but it maintains its full (infinite) Kramers-Moyal
form instead~\cite{Hanggi1981}. As it turns out, a complete solution is
quite demanding for most of the cases.

In this manuscript we study a damped harmonic oscillator subjected to a
random force described by a heterogeneous compound Poisson process which can
be replicated by means of a RLC circuit with random injections of power, or
several other dynamical processes at the so-called complex system level.
Moreover, such kind of problems often conduce to the emergence of stochastic
resonance~\cite{1981_JPA_14_L453,2009_JCP_130_234109} that is found in so
diverse problems spanning from neuroscience and Parkinson's disease~\cite%
{2009_brain_132_2139} to micromechano-electronics~\cite{2006_PRB_73_172302}.
Although this specific system (and its variants) has been studied by
different authors~\cite{livro_vankampen}, in this manuscript we survey the
problem assuming a very fundamental and different approach. Explicitly, we
make direct averages over the noise for different quantities of interest, in
the reciprocal space-time (Fourier-Laplace) domain, by a method that can be
applied to the treatment of coloured noise~\cite{2006_PhysA_365_289},
multiple types of noise~\cite{2008_PRE_77_011103}, thermal conductance
problems~\cite{2009_PRE_79_051116}, or work fluctuation theorems for small
mechanical models~\cite{2010_PRE_82_021112}. Namely, within the Stratonovich
definition, we first focus our efforts on the evaluation of the steady state
probability density function (PDF). Afterwards, we study the evolution of
the injected (dissipated) power into (out of) the system as well as the
total energy of the system and the emergence of stochastic resonance.

\section{\protect\bigskip Exactly solvable model}

Our system can be described by two coupled stochastic differential
equations:
\begin{equation}
\dot{x}(t)=v(t),  \label{01}
\end{equation}%
\begin{equation}
M\,\dot{v}(t)=-k_{0}x(t)-\gamma v(t)+\eta (t),  \label{02}
\end{equation}%
where $\eta $ is a compound heterogeneous Poisson noise,
\begin{equation}
\eta \left( t\right) =\sum\nolimits_{\ell }\Phi \left( t\right) \,\delta
\left( t-t_{\ell }\right) ,  \label{noise}
\end{equation}%
for which the shots (events) occur at a rate $\lambda \left( t\right) $.
Specifically, we will study the following time dependency,
\begin{equation}
\lambda (t)=\lambda _{0}\left[ 1+A\cos (\omega \,t)\right] ,\qquad \left(
0\leq A<1\right) .  \label{lambda1}
\end{equation}%
The case $A=0$ yields the standard homogeneous case. We have opted for a
sinusoidal dependence of the rate $\lambda \left( t\right) $ because of its
ubiquity in a wide range of phenomena~\cite{livro_poisson}.

The Poisson process, $\mathbb{P}$, is a continuous-time stochastic process
belonging to the class of independently distributed random variables whose
generating function between $t$ and $t+\Delta t$ is%
\begin{equation}
\mathcal{G}_{t,t+\Delta t}\left( z\right) =\exp \left[ \left( z-1\right)
\int_{t}^{t+\Delta t}\lambda \left( t^{\prime }\right) \,dt^{\prime }\right]
,  \label{generatingfunction}
\end{equation}%
which is defined by the rate of events, $\lambda \left( t\right) $, such
that,%
\begin{equation}
\left\langle n\left( t\right) \right\rangle =\int_{0}^{t}\lambda \left(
t^{\prime }\right) \,dt^{\prime },  \label{averagekicks}
\end{equation}%
(we use the notation $\left\langle \left( \ldots \right) \right\rangle $ to
represent averages over samples whereas $\overline{\left( \ldots \right) }$
symbolise averages over time). According to its definition, it is only
possible to have a single event at instant $t$. Therefore, from Eq.~(\ref%
{generatingfunction}) and considering the limit $\Delta t\rightarrow 0$, one
can compute the probability that an event occurs is given by $%
\int_{t}^{t+\Delta t}\lambda \left( t\right) \,dt$, while $%
1-\int_{t}^{t+\Delta t}\lambda \left( t\right) \,dt$ is the probability of
zero events. Again, from Eq.~(\ref{generatingfunction}), we can analyse the
probability distribution of the inter-arrival time PDF, $\mathcal{T}\left(
\tau \right) $. The probability that having an event at time $t$ the next
event occurs at $t+\tau $ is given by,%
\begin{equation}
p\left( \tau |t\right) =\frac{d}{d\tau }\left( 1-\exp \left[
-\int_{t}^{t+\tau }\lambda \left( t^{\prime }\right) \,dt^{\prime }\right]
\right) .  \label{pinter}
\end{equation}%
For a heterogeneous Poisson process, one must still pay attention that each
instant has a different weight in the calculation, exactly because of the
time dependence of $\lambda $. This weight, $f\left( t\right) $, is defined
by%
\begin{equation}
f\left( t\right) =\lambda \left( t\right) /\int_{0}^{T}\lambda \left(
t^{\prime }\right) \,dt^{\prime },  \label{fprob}
\end{equation}%
for a heterogeneous Poisson process taking place in the time interval from $0
$ to $T $. Thence, combining Eqs.~(\ref{pinter})~and~(\ref{fprob}) and
integrating over time one obtains,
\begin{equation}
\mathcal{T}\left( \tau \right) =\int_{0}^{T-\tau }f\left( t\right) \,p\left(
\tau |t\right) \,dt.  \label{pdftempo}
\end{equation}%
Concerning the amplitude, $\Phi $, despite being possible to consider
several distribution functions, we will restrict our study to the classical
exponential probability density function for white shot-noises,
\[
P\left( \Phi \right) =\bar{\Phi}^{-1}\exp \left[ -\frac{\Phi }{\bar{\Phi}}%
\right] ,
\]%
whose $n$-th order raw moment is, $\overline{\Phi ^{n}}=n!\,\bar{\Phi}^{n}$.

Going back to Eq.~(\ref{noise}), we can define the Poisson patch, $I$,
between $t$ and $t+\tau ,$%
\[
I\left( \tau \right) \equiv \mathbb{P}\left( t+\tau \right) -\mathbb{P}%
\left( t\right) =\int_{t}^{t+\tau }d\mathbb{P}=\int_{t}^{t+\tau }\eta \left(
t^{\prime }\right) \,dt^{\prime },
\]%
from which we can set out,

\[
\int dI\left( \tau ^{\prime }\right) =\int d\mathbb{P}=\int \eta \left(
t^{\prime }\right) \,\,dt^{\prime },
\]%
where we have omitted the time dependence of the patch for the sake of
simplicity. The average of the Poisson patch is,%
\[
\left\langle I\left( \tau \right) \right\rangle _{c}=\left\langle I\left(
\tau \right) \right\rangle =\bar{\Phi}\int_{t}^{t+\tau }\lambda \left(
t^{\prime }\right) \,dt^{\prime }.
\]%
and its covariance,%
\[
I\left( \tau _{1}\right) I\left( \tau _{2}\right) =\int_{t}^{t+\tau
_{1}}\eta \left( t^{\prime }\right) \,dt^{\prime }\int_{t}^{t+\tau _{2}}\eta
\left( t^{\prime \prime }\right) \,dt^{\prime \prime }.
\]%
which yields after averaging,%
\begin{eqnarray*}
\left\langle I\left( \tau _{1}\right) \,I\left( \tau _{2}\right)
\right\rangle _{c} &=&\overline{\Phi ^{2}}\int_{t}^{t+\tau
_{1}}\int_{t}^{t+\tau _{2}}\lambda \left( t^{\prime \prime }\right) \,\delta
\left( t^{\prime \prime }-t^{\prime }\right) \,dt^{\prime \prime
}\,dt^{\prime } \\
&=&\overline{\Phi ^{2}}\int_{t}^{t+\min \left( \tau _{2},\tau _{1}\right)
}\lambda \left( t^{\prime }\right) \,\,dt^{\prime }.
\end{eqnarray*}

These moments can be straightforwardly generalised to,
\[
I\left( t\right) ^{n}=\int_{t}^{t+\tau }dt_{1}\ldots \int_{t}^{t+\tau
}dt_{n}\,\eta (t_{1})\ldots \eta (t_{n}).
\]%
The noise cumulant averages are,
\[
\langle I\left( t\right) ^{n}\rangle _{c}=\int_{t}^{t+\Delta t}dt_{1}\ldots
\int_{t}^{t+\Delta t}dt_{n}\,\langle \eta (t_{1})\ldots \eta (t_{n})\rangle
_{c}=\lambda \left( t\right) \,\overline{\Phi ^{n}}\,\Delta t.
\]%
This implies that the noise cumulant correlations are identified as~\cite%
{Hanggi1981},
\begin{equation}
\langle \eta (t_{1})\ldots \eta (t_{n})\rangle _{c}=\lambda (t_{1})\,%
\overline{\Phi ^{n}}\,\delta (t_{1}-t_{2})\ldots \delta (t_{n-1}-t_{n}).
\label{eq.04}
\end{equation}

Accordingly,%
\begin{eqnarray}
\int_{t}^{t+\Delta t}\prod_{i=1}^{n}dt_{i}\,\langle \eta (t_{1})\ldots \eta
(t_{n})\rangle _{c} &=&\int_{t}^{t+\Delta t}\prod_{i=1}^{n}dt_{i}\,\lambda
_{0}\left( 1+A\cos (\omega \,t_{1})\right) \Phi ^{n}\prod_{j=1}^{n-1}\delta
(t_{j}-t_{j+1})  \nonumber \\
&=&\lambda _{0}\,\overline{\Phi ^{n}}\int_{t}^{t+\Delta t}dt_{1}\left(
1+A\cos (\omega \,t_{1})\right)  \nonumber \\
&=&\lambda _{0}\,\overline{\Phi ^{n}}\left[ \Delta t+2\frac{A}{\omega }\sin
\left( \omega \,\frac{\,\Delta \,t}{2}\right) \,\cos (\omega \,t)\right] .
\end{eqnarray}%
Taking into account the limit, $\Delta t\rightarrow 0$, the previous
expression tends to $\lambda _{0}\,\overline{\Phi ^{n}}\,\Delta t\left(
1+A\cos (\omega \,t)\right) $ and thus,%
\begin{equation}
\fl\int_{t}^{t+\Delta t}dt_{1}\ldots \int_{t}^{t+\Delta t}dt_{n}\,\langle
\eta (t_{1})\ldots \eta (t_{n})\rangle _{c}=\lambda _{0}\,\overline{\Phi ^{n}%
}\left[ 1+A\cos \left( \omega t\right) \right] \,\Delta t=\lambda \left(
t\right) \,\overline{\Phi ^{n}}\,\Delta t.  \label{lambdaintegration1}
\end{equation}%
In the case $n=1$, Eq.~(\ref{lambdaintegration1}) satisfies the relation $%
\langle I\left( t\right) \rangle =\lambda \left( t\right) \,\bar{\Phi}%
\,\Delta t$, as expected.

Throughout this manuscript we employ the Stratonovich representation for the
noise,%
\[
\left\langle \int_{t}^{t+\Delta t}I\,dI\right\rangle _{c}=\left\langle
\int_{t}^{t+\Delta t}\frac{dI^{2}}{2}\right\rangle _{c}=\lambda \left(
t\right) \,\overline{\Phi ^{2}}\,\frac{\Delta t}{2},
\]%
where the effective noise in a interval $dt$ is computed as the average of
the noise at $t$ and $t+dt$.

\subsection{Laplace transformations}

\label{laplacesec}

Taking the Laplace transformations of Eqs.~(\ref{01})~and~(\ref{02}) (with Re%
$(s)>0$) we obtain\footnote{%
In the Laplace transform we have assumed that both the initial position and
the initial velocity are equal to zero. We are interested in asymptotic
effects and the initial memory terms vanish in that limit.},
\begin{equation}
s\,\tilde{x}(s)=\tilde{v}(s).  \label{eq.12}
\end{equation}%
Defining,
\begin{equation}
R(s)\,\equiv \,s^{2}+\frac{\gamma }{M}s+\frac{k_{0}}{M}=(s-\kappa
_{+})(s-\kappa _{-}),
\end{equation}%
we can write,%
\begin{equation}
\tilde{x}(s)=\frac{\tilde{\eta}(s)}{M\,R(s)},  \label{eq.11}
\end{equation}%
the singularities of which (the zeroes of $R(s)$) are located at,%
\begin{equation}
\kappa _{\pm }=-\theta \pm \,i\,\Omega .
\end{equation}%
with $\theta =\frac{\gamma }{2\,M}$, $\omega _{0}^{2}=\frac{k_{0}}{M}$ and $%
\Omega =\sqrt{\omega _{0}^{2}-\theta ^{2}}.$

For the Poisson process with time-dependent rate, Eq.~(\ref{lambda1}), the
Laplace transform of the noise averages yields,
\begin{eqnarray}
\langle \tilde{\eta}(z_{1})\ldots \tilde{\eta}(z_{n})\rangle _{c}
&=&\int_{0}^{\infty }\prod_{i=1}^{n}dt_{i}\,\exp \left\{
-\sum_{l=1}^{n}z_{l}t_{l}\right\} \langle \eta (t_{1})\ldots \eta
(t_{n})\rangle _{c}  \nonumber \\
&=&\lambda _{0}\,\overline{\Phi ^{n}}\int_{0}^{\infty
}\prod_{i=1}^{n}dt_{i}\,\delta (t_{1}-t_{2})\ldots \delta (t_{n-1}-t_{n})
\nonumber  \label{eq.05} \\
&&\times \left[ 1+A\cos (\omega \,t_{1})\right] \exp \left\{
-\sum_{l=1}^{n}z_{l}t_{l}\right\} ,
\end{eqnarray}%
for which we can separate out its homogeneous and heterogeneous parts. For
the former, we obtain,
\begin{eqnarray}
\mathcal{I}_{1} &=&\lambda _{0}\,\overline{\Phi ^{n}}\int_{0}^{\infty
}dt_{1}\,\exp \left\{ -t_{1}\sum_{l=1}^{n}z_{l}\right\}   \nonumber \\
&=&\frac{\lambda _{0}\,\overline{\Phi ^{n}}}{\sum_{l=1}^{n}z_{l}},
\end{eqnarray}%
whereas the latter is given by,
\begin{eqnarray}
\mathcal{I}_{2} &=&\lambda _{0}\,\overline{\Phi ^{n}}\,A\int_{0}^{\infty
}dt_{1}\,\cos (\omega \,t_{1})\exp \left\{ -t_{1}\sum_{l=1}^{n}z_{l}\right\}
\nonumber \\
&=&\lambda _{0}\,\overline{\Phi ^{n}}\frac{A}{2}\left( \frac{1}{%
\sum_{l=1}^{n}z_{l}-\mathrm{i}\,\omega }+\frac{1}{\sum_{l=1}^{n}z_{l}+%
\mathrm{i}\,\omega }\right) .  \label{lagragecoseno}
\end{eqnarray}

The Laplace transform for the noise cumulants is thus given by:
\begin{equation}
\fl\left\langle \tilde{\eta}(z_{1})\ldots \tilde{\eta}(z_{n})\right\rangle
_{c}=\mathcal{I}_{1}+\mathcal{I}_{2}.
\end{equation}

\section{\protect\bigskip Averaged steady state}

Instead of equilibrium conditions, a periodically forced system reaches a
periodically driven state that is characterised by periodic variations of
the averages and cumulants of its variables~\cite{1996_PLA_215_26}. This
behaviour can now be explicitly obtained by the method previously mentioned~%
\cite{2006_PhysA_365_289}, that we shall use to study the Poisson process.
However, taking the time average for the distribution function does make
sense given that important quantities, such as the injected and dissipated
energies, are well understood when represented by their time averaged
values. In the following, we develop the techniques needed for the exact
solution of Eqs.~(\ref{01})~and~(\ref{02}) at long times. Our results are
comparable with those obtained from the analysis of a single (and
sufficiently long) run of the process in which all the values of the
observable are treated as equally distributed.

We average the probability distribution over time to obtain the cumulant
expansion~\cite{livro_vankampen,2006_PhysA_365_289,2008_PRE_77_011103}
\begin{eqnarray*}
\fl p_{ss}(x,v) &=&\sum_{n,m=0}^{\infty }\int_{-\infty }^{+\infty }\frac{dQ}{%
2\pi }\,\frac{dP}{2\pi }\,e^{\mathrm{i\,}\left( Q\,x+P\,v\right) }\frac{(-%
\mathrm{i\,}Q)^{n}}{n!}\frac{(-\mathrm{i\,}P)^{m}}{m!}\,\overline{ \left\langle
x^{n}v^{m}\right\rangle} \\
&=&\int_{-\infty }^{+\infty }\frac{dQ}{2\pi }\,\frac{dP}{2\pi }\,e^{\mathrm{%
\ i\,}\left( Q\,x+P\,v\right) }\,\exp \left\{ \sum_{n,m=0:(m+n>0)}^{\infty \,}\frac{%
(-\mathrm{i\,}Q)^{n}}{n!}\frac{(-\mathrm{i\,}P)^{m}}{m!} \overline{\left\langle
x^{n}v^{m}\right\rangle _{c}}\right\} .
\end{eqnarray*}%

Let us work out the exact form of the cumulants in terms of the Laplace
transforms of the noise. Applying our definition for computing averages we
have,
\begin{eqnarray*}
\fl\overline{\left\langle x^{n}v^{m}\right\rangle _{c}} &=&\lim_{z%
\rightarrow 0}z\int_{0}^{\infty }dt\,e^{-zt}\,\,\langle
x^{n}(t)\,v^{m}(t)\rangle _{c}\, \\
&=&\lim_{z\rightarrow 0}\lim_{\epsilon \rightarrow 0}\int_{-\infty
}^{+\infty }\prod_{h=1}^{n}\frac{dq_{h}}{2\pi }\int_{-\infty }^{+\infty
}\prod_{j=1}^{m}\frac{dp_{j}}{2\pi }\frac{z}{z-\left[ \sum_{h=1}^{n}(\mathrm{%
i\,}q_{h}+\epsilon )+\sum_{j=1}^{m}(\mathrm{i\,}p_{j}+\epsilon )\right] } \\
&&\times \prod_{h=1}^{n}\frac{1}{\left[ M\,R(\mathrm{i\,}q_{h}+\epsilon )%
\right] }\,\prod_{j=1}^{m}\frac{(\mathrm{i\,}p_{j}+\epsilon )}{\left[ M\,R(%
\mathrm{i\,}p_{j}+\epsilon )\right] }\left\langle \prod_{h=1}^{n}\tilde{\eta}%
(\mathrm{i\,}q_{h}+\epsilon )\prod_{j=1}^{m}\tilde{\eta}(\mathrm{i\,}%
p_{j}+\epsilon )\right\rangle _{c},
\end{eqnarray*}%
where the integration path for the variables in complex space is the same as
in Refs.~\cite{2006_PhysA_365_289,2008_PRE_77_011103}.

\section{General steady state distribution}

Following the previous section, the Poisson steady state distribution
exactly yields,
\begin{equation}
\fl p_{ss}(x,v)=\int_{-\infty }^{+\infty }\frac{dQ}{2\pi }\frac{dP}{2\pi }%
e^{iQx+iPv}\exp \left\{ \sum_{n+m=0:(n+m>0)}^{\infty }\frac{(-\mathrm{i\,}Q)^{n}}{n!}%
\frac{(-\mathrm{i\,}P)^{m}}{m!}\,\mathcal{P}_{n,m}\right\} .  \label{Pnm}
\end{equation}

We may then split Eq.~(\ref{Pnm}) into two parts: term $\mathcal{P}^{\left(
1\right) }$ which arises from the time-independent contribution,

\begin{equation}
\fl\mathcal{P}_{n,m}^{\left( 1\right) }=\int_{-\infty }^{+\infty
}\prod_{h=1}^{m+n-1}\frac{dp_{h}}{2\pi }\frac{\prod_{h=1}^{m}(\mathrm{i\,}%
p_{h}+\epsilon )}{\prod_{h=1}^{m+n-1}R(\mathrm{i\,}p_{h}+\epsilon )}\,\,%
\frac{1}{R\left( -\sum_{h=1}^{m+n-1}(\mathrm{i}\,p_{h}+\epsilon )\right) },
\label{parteHomo}
\end{equation}%
and the remaining terms arising from the periodic forcing term,
\[
\fl\mathcal{P}_{n,m}^{\left( \pm \right) }=\lim_{z\rightarrow 0}\frac{%
\lambda _{0}\,A}{2}\int_{-\infty }^{+\infty }\prod_{h=1}^{n+m-1}\frac{dp_{h}%
}{2\pi }\frac{z}{z\pm \mathrm{i}\,\omega }\frac{\prod_{h=1}^{m}(\mathrm{i\,}%
p_{h}+\epsilon )}{\prod_{h=1}^{n+m-1}R(\mathrm{i\,}p_{h}+\epsilon )}\frac{1}{%
R\left( \pm \mathrm{i}\,\omega -\sum_{h=1}^{n+m}(\mathrm{i\,}p_{h}+\epsilon
)\right) }.
\]%
Unrolling Eq.~(\ref{parteHomo}), we can determine the different
contributions that are allowed to emerge by{\ taking into consideration the
different powers of }$x$ and $v$ in the cumulant. Accordingly, term $\Psi
_{1}$ represents cumulants of zeroth order in the velocity, $\left\langle
x^{n}\right\rangle _{c}$,
\begin{equation}
\fl\Psi _{1x}=\sum\limits_{j=0}^{n-1}\left(
\begin{array}{c}
n-1 \\
j%
\end{array}%
\right) \frac{\left( -1\right) ^{n-j-1}}{\left[ 1,-1\right] ^{n-1}\left[
j\,,\left( n-j\right) \,\right] \left[ \left( j+1\right) ,\left(
n-j-1\right) \,\right] },  \label{psi1x}
\end{equation}%
and in the position, $\left\langle v^{m}\right\rangle _{c}$,%
\begin{equation}
\fl\Psi _{1v}=\sum\limits_{j=0}^{m-1}\left(
\begin{array}{c}
m-1 \\
j%
\end{array}%
\right) \frac{\mathrm{i}^{\,m}\left( -1\right)
^{m-j}\,k_{+}^{\,j}\,k_{-}^{m-j-1}\left[ j\,,\left( m-j-1\right) \,\right] }{%
\left[ 1,-1\right] ^{m-1}\left[ j\,,\left( m-j\right) \,\right] \left[
\left( j+1\right) ,\left( m-j-1\right) \,\right] },  \label{psi1v}
\end{equation}%
Term $\Psi _{2}$ describes cumulants which are such as $\left\langle
x^{n-1}v\right\rangle _{c}$,%
\begin{equation}
\fl\Psi _{2}=\sum\limits_{j=0}^{n-2}\left(
\begin{array}{c}
n-2 \\
j%
\end{array}%
\right) \frac{\mathrm{i\,}\left( -1\right) ^{n-j-1}}{\left[ 1,-1\right]
^{n-2}}\frac{\left[ j,\left( n-j-2\right) \right] }{\left[ j,\left(
n-j-1\right) \,\right] \left[ \left( j+1\right) \,,\left( n-j-2\right) %
\right] },
\end{equation}%
and last term {$\Psi $}$_{{3}}$ represents the remaining combinations of
powers of $x^{l}$ and $v^{m}$ with $l+m=n$ and $l\geq 2$,{%
\begin{eqnarray}
\fl\Psi _{3}\left( m\right)
&=&\sum\limits_{j=0}^{m}\sum_{l=0}^{n-m-1}\left(
\begin{array}{c}
m \\
j%
\end{array}%
\right) \left(
\begin{array}{c}
n-m-1 \\
l%
\end{array}%
\right) \frac{\left( -1\right) ^{n-m+j}}{\left[ 1,-1\right] ^{n-m-1}}\frac{%
\mathrm{i}^{m+1}\,k_{+}^{j}\,k_{-}^{m-j}\,\left[ j\,,\left( m-j\right) \,%
\right] }{\left[ 1,-1\right] ^{m}}\times   \nonumber \\
&\times &\frac{1}{\left[ \left( j+l+1\right) \,,\left( n-m-j-l-1\right) \,%
\right] \left[ \left( j+l\right) \,,\left( n-m-j-l\right) \,\right] },
\end{eqnarray}%
in which we have the used the following curtailed notation,
\[
\left( a\,\kappa _{+}+b\,\kappa _{-}\right) \equiv \left[ a,b\right] .
\]%
} Therefore, bearing in mind the last definition of the cumulant part of the
probability distribution, Eq.~(\ref{parteHomo}), we can write it as:
\begin{eqnarray}
\hspace{-2.5cm}p_{ss}(x,v) &=&\mathcal{F}_{x,v}\left[ \exp \left\{
\sum_{n,m=0:(m+n>0)}^{\infty }\lambda _{0}\,(n+m)!\,\frac{Q^{n}}{n!}\frac{P^{m}}{m!}%
\,\left( \mathrm{i\,}\bar{\Phi}\right) ^{n+m}\times \right. \right.
\nonumber \\
&&\left. \left. \left( \Psi _{1x}\,\delta _{m,0}+\Psi _{1v}\,\delta
_{n,0}+\Psi _{2}\,\delta _{m,1}+\sum\limits_{m=2}^{n+m-1}\Psi _{3}\left(
m\right) \right) \right\} \right] ,  \label{p0eq}
\end{eqnarray}%
where $\mathcal{F}_{\left( x,v\right) }\left[ f\left( Q,P\right) \right] $
represents the 2-dimensional Fourier Transform into position-velocity real
space, $\left( x,v\right) $.

Owing to the factor $z/\left( z\mp \mathrm{i}\,\omega \right) $ in terms $%
\mathcal{P}^{\left( \pm \right) }$ we are able to verify that after
performing integrations following the appropriate contour, the $z$ terms
hold on up to the end, so that when we finally compute the limit of $%
z\rightarrow 0$ both contributions vanish. Therefore, for this type of
heterogeneity, the steady state distribution bears out the same result as
the homogeneous Poisson process with constant rate $\lambda _{0}$. This is
quite understandable since we are making a long time average in which the
contribution of the periods whose rate is larger than $\lambda _{0}$ kills
off the contribution arising from the periods in which the rate is smaller
than $\lambda _{0}$ because of the symmetry of the rate around $\lambda _{0}$%
.

Regarding the marginal steady state distributions,
\begin{equation}
p_{ss}\left( x\right) =\int p_{ss}(x,v)\,dv,  \label{peqx1}
\end{equation}%
and%
\begin{equation}
p_{ss}\left( v\right) =\int p_{ss}(x,v)\,dx,  \label{peqv1}
\end{equation}%
we start with the probability distribution of the position and following our
procedure we obtain,%
\begin{equation}
p_{ss}\left( x\right) =\mathcal{F}_{x}\left[ \exp \left\{ \sum_{n>0}^{\infty
}\lambda _{0}\mathrm{\,}Q^{n}\,\bar{\Phi}^{n}\,\Psi _{1x}\right\} \right] ,
\label{peqx2}
\end{equation}%
whence we can identify the cumulants,%
\begin{equation}
\kappa _{n}\equiv \overline{\left\langle x^{n}\right\rangle _{c}}=n!\lambda
_{0}\,\left( \mathrm{i\,}\bar{\Phi}\right) ^{n}\,\Psi _{1x}.
\end{equation}%
Using the property of Pascal triangles,%
\begin{equation}
\left(
\begin{array}{c}
n \\
j%
\end{array}%
\right) =\left(
\begin{array}{c}
n-1 \\
j-1%
\end{array}%
\right) +\left(
\begin{array}{c}
n-1 \\
j%
\end{array}%
\right) ,
\end{equation}%
we can write the sum in Eq.~(\ref{psi1x}) over the index $j$ as,

\begin{equation}
\sum_{j=0}^{n-1}\left[ \ldots \right] =\,n!\,\left[ 1,-1\right] ^{n-1}\frac{1%
}{D_{n}},
\end{equation}%
where,%
\begin{equation}
D_{n}=\prod_{j=0}^{n}\left[ j,n-j\right] .  \label{Dn_eq}
\end{equation}%
Accordingly, the cumulants of $p_{ss}(x)$ are%
\begin{equation}
\kappa _{n}^{\left( x\right) }=\lambda _{0}\,\,\left( \frac{\bar{\Phi}}{M}%
\right) ^{n}(-1)^{n-1}\frac{\left( n!\right) ^{2}}{D_{n}}\qquad \left( n\geq
1\right) .  \label{rawmoments}
\end{equation}%
Allowing for Eq.~(\ref{rawmoments}), we explicit the average,%
\[
\overline{\left\langle x\right\rangle }=\bar{\Phi}\,\frac{\lambda _{0}}{k_{0}%
},
\]%
and the second-order moment,
\[
\overline{\left\langle x^{2}\right\rangle }-\overline{\left\langle
x\right\rangle }^{2}=\bar{\Phi}^{2}\frac{\lambda _{0}}{\gamma \,k_{0}}.
\]

We have implemented a computational procedure to numerically compute the
probability density function of the position at the steady state. Our
exhibited numerical results were obtained for different values of $M$, $k_{0}
$ and $\gamma $ and fixed values of $\lambda _{0}=10$, $\bar{\Phi}=1$ and $%
\omega =\pi $ following the implementation described in \ref{appA}. From
cases A and B (see values in the figure caption), we can understand that the
mass does not impact in both the average and standard deviation and that B
and C tally, although each is based on a homogeneous and a heterogeneous
process, respectively. Comparing cases A, B and C we verify that the lighter
the particle, the skewer the distribution $p_{ss}\left( x\right) $. The
results of cases B, C and E help us show that $\gamma $ does not affect the
average and finally case F sketches the influence of $k_{0}$. Case D permit
us to follow the dependence on the mass. Lighter particles have a skewer
distribution. The results are clearly different if we consider a symmetric
noise with $\bar{\Phi}=0$. As we will see in the next section, the positive
total injection of power associated with this case is replaced by a zero
average injection of power related to the steady state average position
being zero.

\begin{figure}[tbh]
\begin{center}
\includegraphics[width=0.60\columnwidth,angle=0]{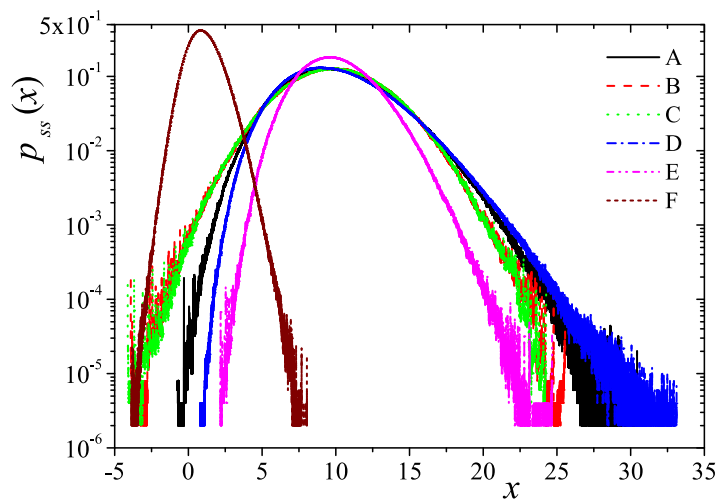}
\end{center}
\caption{Numerically obtained probability density function $p_{ss}\left(
x\right) $ vs position $x$ for various cases with $\protect\lambda _{0}=10$,
$\bar{\Phi}=1$ and the noise defined by Eq.~(\protect\ref{lambda1}) with $%
\protect\omega =\protect\pi $. Following the legend in the figure we have
the respective cases, A: $M=1,k_{0}=1,\protect\gamma =1,A=0$, B: $%
M=10,k_{0}=1,\protect\gamma =1,A=0$, C: $M=10,k_{0}=1,\protect\gamma %
=1,A=1/2 $, D: $M=0.1,k_{0}=1,\protect\gamma =1,A=0$, E: $M=1,k_{0}=1,%
\protect\gamma =2,A=0$ and F: $M=1,k_{0}=10,\protect\gamma =1,A=0$.}
\label{fig-peqx}
\end{figure}

The picture is very much the same for the marginal distribution of the
velocity. Namely, from Eq.~(\ref{peqv1}) we get,
\[
p_{ss}\left( v\right) =\mathcal{F}_{v}\left[ \exp \left\{ \sum_{m>0}^{\infty
}\lambda _{0}\,P^{m}\,\bar{\Phi}^{m}\,\Psi _{1v}\right\} \right] ,
\]%
and consequently the cumulants are,
\begin{equation}
\kappa _{m}^{\left( v\right) }\equiv \overline{\left\langle
v^{\,m}\right\rangle _{c}}=m!\,\lambda _{0}\,\left( \mathrm{i\,}\bar{\Phi}%
\right) ^{m}\,\Psi _{1v}.
\end{equation}

\begin{figure}[tbh]
\begin{center}
\includegraphics[width=0.60\columnwidth,angle=0]{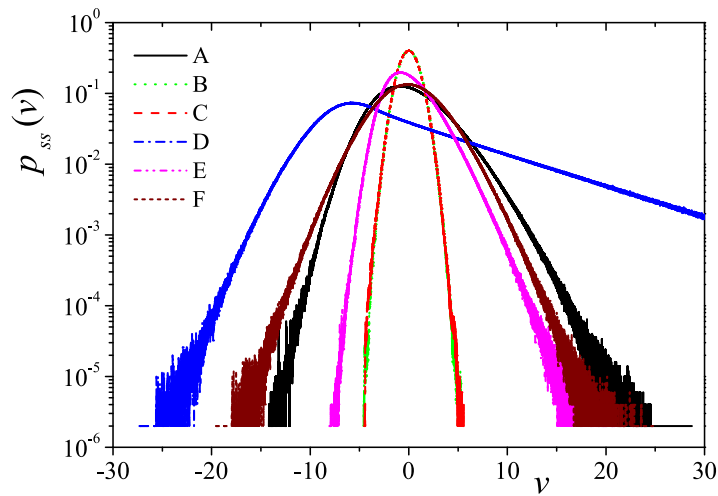}
\end{center}
\caption{Numerically obtained probability density function $p_{ss}\left(
v\right) $ vs scalar velocity $v$ for the same parameters sets of Fig.~%
\protect\ref{fig-peqx}. }
\label{fig-peqv}
\end{figure}

For the cumulants of the marginal velocity distribution, the calculation
turns out much harder and haplessly we have not managed to write it in a
compact form as Eq.~(\ref{peqx2}). Nevertheless, we can still write some of
them explicitly, such as the first,
\[
\kappa _{1}^{v}\equiv \overline{\left\langle v\right\rangle }=0,
\]%
and the second cumulants,
\[
\kappa _{2}^{v}\equiv \overline{\left\langle v^{2}\right\rangle }=\frac{%
\lambda _{0}\,\bar{\Phi}^{2}}{M\,\gamma }.
\]%
In Fig.~\ref{fig-peqv}, we plot the results of $p_{ss}\left( v\right) $ for
the same the numerical implementation of Fig.~\ref{fig-peqx}. Once more, we
can understand the independence of the probability distribution regarding
the amplitude $A$ in Eq.~(\ref{lambda1}). We can also notice the influence
of the mass, lighter particles are more sensitive to the noise and thus, for
the same noise intensity, they achieve larger positive values of the
velocity. Moreover, it is visible that $p_{ss}\left( v\right) $ might be
strongly positively skew for light particles. On the other hand, if we
consider an average over samples the effect of the amplitude $A$ and
frequency $\omega $ of the Poisson noise.

\section{\protect\bigskip \protect\bigskip Injection and dissipation of
Energy}

An interesting element of study, especially in practical applications such
as presented in Ref.~\cite{2006_PRB_73_172302,1981_JPA_14_L453}, concerns
the time evolution of energy (related) quantities. This can be checked by
heeding the fact that variations of energy in an isolated system equals the
total work done by the external forces acting on it, in this case the
fluctuating force $\eta $ and the dissipative one $-\gamma v$:
\begin{eqnarray}
\sum W_{F_{ext,i}} &=&\Delta E_{m}  \label{verificacao} \\
\int \eta \,dx-\gamma \int v\,dx &=&\frac{1}{2}M\,v\left( t\right) ^{2}+%
\frac{1}{2}k_{0}\,x\left( t\right) ^{2}  \nonumber
\end{eqnarray}

The injection of the energy balance can be also analysed by determining the
evolution of the total energy of the particle,%
\begin{equation}
\int_{0}^{\tau }\left[ \eta \left( t\right) \,v\left( t\right) -\gamma
v\left( t\right) ^{2}\right] dt=\frac{1}{2}M\,\left. v\left( t\right)
^{2}\right\vert _{t=0}^{t=\tau }+\frac{1}{2}k_{0}\,\left. x\left( t\right)
^{2}\right\vert _{t=0}^{t=\tau }.  \label{expenerg}
\end{equation}

In what follows, we shall omit transient terms, keeping the more interesting
asymptotic ones whenever possible.

\subsection{\protect\bigskip Energetic considerations}

The average values of $\left[ v\left( t\right) \right] ^{2}$ and $\left[
x\left( t\right) \right] ^{2}$ that we use in the computation of the total
energy can be obtained once more using the Laplace representation,
\begin{eqnarray*}
v\left( t\right) ^{2} &=&\int \int v\left( t_{1}\right) v\left( t_{2}\right)
\,\delta \left( t-t_{1}\right) \,\delta \left( t-t_{2}\right)
\,dt_{1}\,dt_{2} \\
&=&\lim_{\varepsilon \rightarrow 0}\int \int e^{\left( \mathrm{i}\,q_{1}+%
\mathrm{i}\,q_{2}+2\varepsilon \right) t}\,\tilde{v}\left( \mathrm{i}%
\,q_{1}+\varepsilon \right) \tilde{v}\left( \mathrm{i}\,q_{2}+\varepsilon
\right) \,\frac{dq_{1}}{2\,\pi }\,\frac{dq_{2}}{2\,\pi },
\end{eqnarray*}%
whence by averaging and taking into account the second cumulant definition,
\begin{eqnarray*}
\fl\left\langle v\left( t\right) ^{2}\right\rangle -\left\langle v\left(
t\right) \right\rangle ^{2} &=&\lim_{\varepsilon \rightarrow 0}\frac{1}{M^{2}%
}\int \int e^{\left( \mathrm{i}\,q_{1}+\mathrm{i}\,q_{2}+2\varepsilon
\right) t}\,\left( \mathrm{i}\,q_{1}+\varepsilon \right) \left( \mathrm{i}%
\,q_{2}+\varepsilon \right) \,\times \\
&&\frac{\left\langle \tilde{\eta}\left( \mathrm{i}\,q_{1}+\varepsilon
\right) \,\tilde{\eta}\left( \mathrm{i}\,q_{2}+\varepsilon \right)
\right\rangle _{c}}{R\left( \mathrm{i}\,q_{1}+\varepsilon \right) \,R\left(
\mathrm{i}\,q_{2}+\varepsilon \right) }\frac{dq_{1}}{2\,\pi }\frac{dq_{2}}{%
2\,\pi },
\end{eqnarray*}%
and,
\[
\fl\left\langle x\left( t\right) ^{2}\right\rangle -\left\langle x\left(
t\right) \right\rangle ^{2}=\lim_{\varepsilon \rightarrow 0}\frac{1}{M^{2}}%
\int \int e^{\left( \mathrm{i}\,q_{1}+\mathrm{i}\,q_{2}+2\varepsilon \right)
t}\,\frac{\left\langle \tilde{\eta}\left( \mathrm{i}\,q_{1}+\varepsilon
\right) \,\tilde{\eta}\left( \mathrm{i}\,q_{2}+\varepsilon \right)
\right\rangle _{c}}{R\left( \mathrm{i}\,q_{1}+\varepsilon \right) \,R\left(
\mathrm{i}\,q_{2}+\varepsilon \right) }\frac{dq_{1}}{2\,\pi }\,\frac{dq_{2}}{%
2\,\pi },
\]%
where the asymptotic solutions are,
\begin{eqnarray*}
\fl\left\langle v\left( t\right) ^{2}\right\rangle _{asy}-\left\langle
v\left( t\right) \right\rangle _{asy}^{2} &=&{\frac{\lambda _{{0}}{\bar{\Phi}%
}^{2}}{\gamma \,M}}+8\,{\frac{A\left[ 4\,\left( {\omega }^{2}{\theta }%
^{3}+\,\theta \,{\omega _{{0}}}^{4}\right) \cos \left( \omega \,t\right) +\,{%
\omega }^{3}{\theta }^{2}\sin \left( \omega \,t\right) \right] \lambda _{{0}}%
{\bar{\Phi}}^{2}}{{M}^{2}\left( 4\,{\theta }^{2}+{\omega }^{2}\right) \left(
\left( \omega _{0}^{2}-\omega ^{2}\right) ^{2}+4\,{\theta }^{2}{\omega }%
^{2}\right) }}+ \\
&&4\,{\frac{A\left[ 2\,\theta \left( {\omega }^{4}-3\,{\omega _{{0}}}^{2}{%
\omega }^{2}\right) \cos \left( \omega \,t\right) -3\,{\omega _{{0}}}^{2}{%
\omega }^{3}\sin \left( \omega \,t\right) \right] \lambda _{{0}}{\bar{\Phi}}%
^{2}}{{M}^{2}\left( 4\,{\theta }^{2}+{\omega }^{2}\right) \left( \left(
\omega _{0}^{2}-\omega ^{2}\right) ^{2}+4\,{\theta }^{2}{\omega }^{2}\right)
}}+ \\
&&2\,{\frac{A\left( {\omega }^{5}+8\,{\omega _{{0}}}^{4}\omega \,\right)
\,\lambda _{{0}}{\bar{\Phi}}^{2}\sin \left( \omega \,t\right) }{{M}%
^{2}\left( 4\,{\theta }^{2}+{\omega }^{2}\right) \left( \left( \omega
_{0}^{2}-\omega ^{2}\right) ^{2}+4\,{\theta }^{2}{\omega }^{2}\right) },}
\end{eqnarray*}%
and

\begin{eqnarray*}
\fl\left\langle x\left( t\right) ^{2}\right\rangle _{asy}-\left\langle
x\left( t\right) \right\rangle _{asy}^{2} &=&\bar{\Phi}^{2}\frac{\lambda _{0}%
}{\gamma \,k_{0}}+{\frac{4\omega \left( {\omega }^{2}-4\,{\omega _{{0}}}%
^{2}-8\,{\theta }^{2}\right) \,A\,{\bar{\Phi}}^{2}\lambda _{{0}}\sin \left(
\omega \,t\right) }{\left( {\omega }^{4}-8\,{\omega _{{0}}}^{2}{\omega }%
^{2}+16\,{\theta }^{2}{\omega }^{2}+16\,{\omega _{{0}}}^{4}\right) \left( 4\,%
{\theta }^{2}+{\omega }^{2}\right) {M}^{2}}}+ \\
&&{\frac{8\left( 3\,\theta \,{\omega }^{2}-2\,\theta \,{\omega _{{0}}}%
^{2}\right) \,A\,{\bar{\Phi}}^{2}\lambda _{{0}}\cos \left( \omega \,t\right)
\,}{\left( {\omega }^{4}-8\,{\omega _{{0}}}^{2}{\omega }^{2}+16\,{\theta }%
^{2}{\omega }^{2}+16\,{\omega _{{0}}}^{4}\right) \left( 4\,{\theta }^{2}+{%
\omega }^{2}\right) {M}^{2}}}.
\end{eqnarray*}

We must now take into account the square values of $\left\langle v\left(
t\right) \right\rangle_{asy} $ and $\left\langle x\left( t\right)
\right\rangle_{asy} $, which, for all times, are given by,
\begin{eqnarray*}
\fl \left\langle x\left( t\right) \right\rangle &= &\lim_{\varepsilon
\rightarrow 0}\frac{1}{M} \int\,\frac{dq_{1}}{2\,\pi } \, e^{\left( \mathrm{i%
}\,q_{1}+\varepsilon \right) t}\,\frac{\left\langle \tilde{\eta}\left(
\mathrm{i}\,q_{1}+\varepsilon \right) \, \right\rangle }{R\left( \mathrm{i}%
\,q_{1}+\varepsilon \right) } \\
&= &\lim_{\varepsilon \rightarrow 0}\frac{\lambda_0\,\bar{\Phi}}{M} \int\,%
\frac{dq_{1}}{2\,\pi } \, \,\frac{e^{\left( \mathrm{i}\,q_{1}+\varepsilon
\right) t} }{R\left( \mathrm{i}\,q_{1}+\varepsilon \right) } \left\{ \frac{1%
}{\mathrm{i}\,q_{1}+\epsilon }+\frac{A}{2}\left( \frac{1}{\mathrm{i}%
\,q_{1}+\epsilon -\mathrm{i}\,\omega }+\frac{1}{\mathrm{i}\,q_{1}+\epsilon +%
\mathrm{i}\,\omega }\right)\right\},
\end{eqnarray*}
and,
\begin{eqnarray*}
\fl \left\langle v\left( t\right) \right\rangle &= &\lim_{\varepsilon
\rightarrow 0}\frac{\lambda_0\,\bar{\Phi}}{M} \int\,\frac{dq_{1}}{2\,\pi }
\, \,\frac{e^{\left( \mathrm{i}\,q_{1}+\varepsilon \right) t} }{R\left(
\mathrm{i}\,q_{1}+\varepsilon \right) } \left\{ 1+\frac{A}{ 2}\left( \frac{%
\mathrm{i}\,q_{1}+\epsilon }{\mathrm{i}\,q_{1}+\epsilon -\mathrm{i} \,\omega
}+\frac{\mathrm{i}\,q_{1}+\epsilon }{\mathrm{i}\,q_{1}+\epsilon +\mathrm{\ i}%
\,\omega }\right)\right\}.
\end{eqnarray*}

After integrating over the poles above, we obtain the asymptotic behaviour,
\begin{eqnarray*}
\left\langle v\left( t\right) \right\rangle _{asy} &=&{\frac{\,\,A\left[
\left( {\omega }^{2}-{\ \omega _{{0}}}^{2}\right) \sin \left( \omega
\,t\right) +2\,\omega \,\theta \,\cos \left( \omega \,t\right) \right]
\omega \,\lambda _{{0}}\bar{\Phi}}{M\,\,\left[ \left( \omega _{0}^{2}-\omega
^{2}\right) ^{2}+4\,{\theta }^{2}{\ \omega }^{2}\right] },} \\
\left\langle x\left( t\right) \right\rangle _{asy} &=&{\frac{\bar{\Phi}%
\,\lambda _{{0}}}{{\omega _{{0}}^{2}\,}M}}-{\frac{A\,\left[ \left( {\omega }%
^{2}-{\omega _{{0}}}^{2}\right) \cos \left( \omega \,t\right) -2\,\theta
\,\omega \,\sin \left( \omega \,t\right) \right] \lambda _{{0}}\bar{\Phi}}{%
M\,\left[ \left( \omega _{0}^{2}-\omega ^{2}\right) ^{2}+4\,{\theta }^{2}{\
\omega }^{2}\right] }},
\end{eqnarray*}%
where it can be easily seen that,%
\[
\left\langle v\left( t\right) \right\rangle _{asy}=\frac{d}{dt}%
\,\left\langle x\left( t\right) \right\rangle _{asy}.
\]

\begin{figure}[tbh]
\begin{center}
\includegraphics[width=0.60\columnwidth,angle=0]{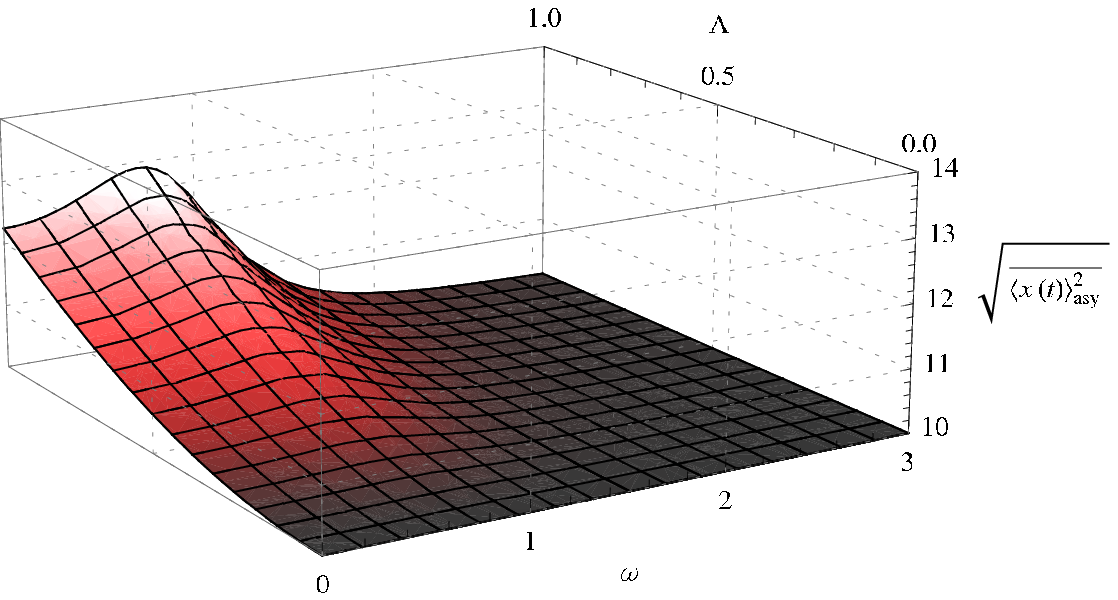} %
\includegraphics[width=0.60\columnwidth,angle=0]{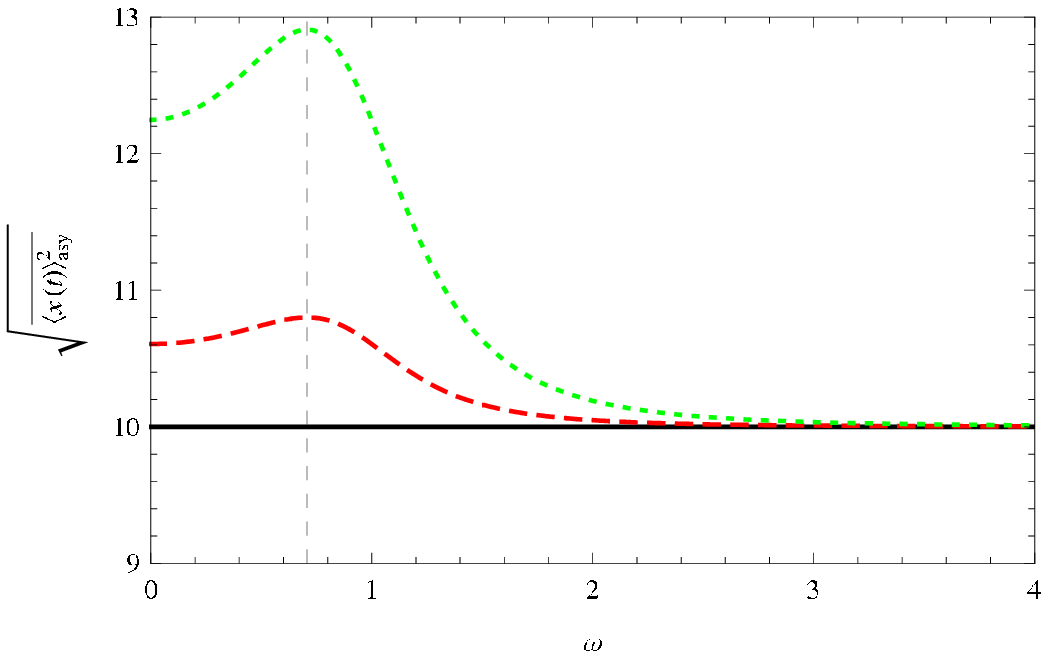}
\end{center}
\caption{Upper panel: $\protect\sqrt{\overline{\left\langle x\left( t\right)
\right\rangle _{asy}^{2}}}$ vs amplitude of the heterogeneous part, $A$, and
frequency, $\protect\omega $, with $M=1$, $\protect\gamma =1$, $k_{0}=1$, $%
\protect\lambda _{0}=10$, $\bar{\Phi}=1$. The maximum, characterising a
stochastic resonance phenomenon, occurs at $\protect\omega _{\mathrm{res}}=%
\protect\sqrt{\protect\omega _{0}^{2}-2\,\protect\theta ^{2}}$ that defines
the grey dashed line in the lower panel containing the plane cuts for $A=0$
(black line), $A=\frac{1}{2}$ (dashed red line) and $A=1$ (dotted green
line). For this case the maximum occurs at $1/\protect\sqrt{2}$.}
\label{resonance}
\end{figure}

We observe that the results above are those expected when we interpret the
oscillating rate Poisson process as a periodic forcing acting upon a damped
harmonic oscillator. As expected, the amplitude of motion shows the typical
resonant behaviour. In order to illustrate this behaviour, we plot in Fig.~%
\ref{resonance} the quantity,
\begin{equation}
\sqrt{\overline{\left\langle x\left( t\right) \right\rangle _{asy}^{2}}}=%
\frac{\bar{\Phi}\,\lambda _{{0}}}{{\omega _{{0}}^{2}}\,M}\sqrt{1+A^{2}\frac{%
\omega _{0}^{4}}{2\left[ 4\theta ^{2}\omega ^{2}+\left( \omega
_{0}^{2}-\omega ^{2}\right) ^{2}\right] }},
\end{equation}
as a function of the amplitude of the oscillating contribution, $A$, and the
frequency of these oscillations, $\omega $. In the both planels is evident
the emergence of a maximum at a frequency of the heterogeneous Poisson rate $%
\lambda \left( t\right) $ equal to $\sqrt{\omega _{0}^{2}-2\theta ^{2}}$.

Adding all the terms and taking the limit $t\rightarrow \infty $, we obtain
the equilibrium energy of the system. The energy is composed of an
oscillating term, with time zero average, and a constant term $E_{M}^{c}$:
\begin{eqnarray}
E_{M}^{c} &=&\frac{1}{2}M\overline{\left\langle v\left( t\right)
^{2}\right\rangle _{asy}}+\frac{1}{2}k_{0}\,\overline{\left\langle x\left(
t\right) ^{2}\right\rangle _{asy}}  \nonumber \\
&=&\lambda _{0}\frac{\bar{\Phi}^{2}}{\gamma }+\,{\frac{{\lambda _{{0}}}^{2}{%
\bar{\Phi}}^{2}}{2\,M\,{\omega _{{0}}^{2}}}}+{\frac{{\lambda _{{0}}}^{2}{%
\bar{\Phi}}^{2}{A}^{2}\left( {\omega }^{2}+{\ \omega _{{0}}}^{2}\right) }{%
4\,M\left[ \left( \omega _{0}^{2}-\omega ^{2}\right) ^{2}+4\,{\theta }^{2}{\
\omega }^{2}\right] }}.  \label{energiatotal}
\end{eqnarray}%
It is worth mentioning that in this case we have made explicit the
asymptotic time dependence so that our averages are computed over samples
and not over time in a single sample as we have made in the previous section.

\subsection{Power considerations}

Going back to Eq.~(\ref{expenerg}), we can define the two following
quantities,%
\begin{equation}
J_{I}=v(t)\,\eta (t),
\end{equation}%
and
\begin{equation}
J_{D}=-\gamma \,v^{2}(t).
\end{equation}%
Physically, both rates, $J_{I}$ and $J_{D}$, constitute changes of energy
due to the interactions with the thermal bath. Within this context, it is
particularly important the study of the cumulative changes of energy in the
system up to a time $t=\tau $, namely the injected total,
\begin{equation}
J_{IT}=\int_{0}^{\tau }dt\,v(t)\,\eta (t),  \label{JIT}
\end{equation}%
and the dissipated total,
\begin{equation}
J_{DT}=-\gamma \,\int_{0}^{\tau }dt\,\,v^{2}(t).  \label{JDT}
\end{equation}%
in which we will apply the same Laplace transform operation in order to
better handle the noise averages \cite%
{2006_PhysA_365_289,2008_PRE_77_011103,2009_PRE_79_051116,2010_PRE_82_021112}%
. The dissipation of energy flux can be written as
\begin{eqnarray*}
\fl J_{DT} &=&-\gamma \int_{0}^{\tau }dt\,\int_{0}^{\infty }dt_{1}\,\delta
(t-t_{1})\,\int_{0}^{\infty }dt_{2}\,\delta (t-t_{2})\,v(t_{1})\,\eta
(t_{2}), \\
&=&\lim_{\epsilon \rightarrow 0}\int_{-\infty }^{\infty }\frac{dq_{1}}{2\pi }%
\,\int_{-\infty }^{\infty }\frac{dq_{2}}{2\pi }\,\frac{e^{(\mathrm{i}\,q_{1}+%
\mathrm{i}\,q_{2}+2\epsilon )\tau }-1}{(\mathrm{i}\,q_{1}+\mathrm{i}%
\,q_{2}+2\epsilon )}\,\tilde{v}(\mathrm{i}\,q_{1}+\epsilon )\,\tilde{v}(%
\mathrm{i}\,q_{2}+\epsilon ), \\
&=&\lim_{\epsilon \rightarrow 0}\int_{-\infty }^{\infty }\frac{dq_{1}}{2\pi }%
\,\int_{-\infty }^{\infty }\frac{dq_{2}}{2\pi }\,\frac{e^{(\mathrm{i}\,q_{1}+%
\mathrm{i}\,q_{2}+2\epsilon )\tau }-1}{(\mathrm{i}\,q_{1}+\mathrm{i}%
\,q_{2}+2\epsilon )}\,\times \\
&&\left[ \frac{(\mathrm{i}\,q_{1}+\epsilon )}{M\,R(\mathrm{i}%
\,q_{1}+\epsilon )}\frac{(\mathrm{i}\,q_{2}+\epsilon )}{M\,R(\mathrm{i}%
\,q_{2}+\epsilon )}\tilde{\eta}(\mathrm{i}\,q_{1}+\epsilon )\tilde{\eta}(%
\mathrm{i}\,q_{2}+\epsilon )\right] ,
\end{eqnarray*}
which becomes, after taking the thermal average,
\begin{eqnarray*}
\fl\left\langle J_{DT}\right\rangle &=&\left( -\frac{2\gamma \lambda _{0}\,{%
\bar{\Phi}}^{2}}{M^{2}}\right) \lim_{\epsilon \rightarrow 0}\int_{-\infty
}^{\infty }\frac{dq_{1}}{2\pi }\,\int_{-\infty }^{\infty }\frac{dq_{2}}{2\pi
}\,\frac{e^{(\mathrm{i}\,q_{1}+\mathrm{i}\,q_{2}+2\epsilon )\tau }-1}{(%
\mathrm{i}\,q_{1}+\mathrm{i}\,q_{2}+2\epsilon )}\,\frac{(\mathrm{i}%
\,q_{1}+\epsilon )}{R(\mathrm{i}\,q_{1}+\epsilon )}\frac{(\mathrm{i}%
\,q_{2}+\epsilon )}{R(\mathrm{i}\,q_{2}+\epsilon )}\times \\
&&\left[ \frac{1}{\mathrm{i}\,q_{1}+\mathrm{i}\,q_{2}+2\,\epsilon }+\frac{A}{%
2}\left( \frac{1}{\mathrm{i}\,q_{1}+\mathrm{i}\,q_{2}+2\,\epsilon -\mathrm{i}%
\,\omega }+\frac{1}{\mathrm{i}\,q_{1}+\mathrm{i}\,q_{2}+2\,\epsilon +\mathrm{%
i}\,\omega }\right) \right. + \\
&&\frac{\lambda _{0}}{2}\,\left\{ \frac{1}{\mathrm{i}\,q_{1}+\epsilon }+%
\frac{A}{2}\left( \frac{1}{\mathrm{i}\,q_{1}+\epsilon -\mathrm{i}\,\omega }+%
\frac{1}{\mathrm{i}\,q_{1}+\epsilon +\mathrm{i}\,\omega }\right) \right\}
\times \\
&&\left. \left\{ \frac{1}{\mathrm{i}\,q_{2}+2\,\epsilon }+\frac{A}{2}\left(
\frac{1}{\mathrm{i}\,q_{2}+\epsilon -\mathrm{i}\,\omega }+\frac{1}{\mathrm{i}%
\,q_{2}+\epsilon +\mathrm{i}\,\omega }\right) \right\} \right] \\
&\equiv &J_{DT0}(\tau )+J_{D,osc}(\tau ).
\end{eqnarray*}

Under the condition $\tau \theta \gg 1$ (transient terms are negligible), we
can explicitly write the results for the integration above as a sum of three
contributions: a term proportional to $\tau $,
\[
\fl J_{DT0}(\tau )=-\left( {\frac{\,{\bar{\Phi}}^{2}{\lambda _{{0}}}}{M}}+{%
\frac{{A}^{2}{\omega }^{2}{\bar{\Phi}}^{2}{\lambda _{{0}}^{2}\,}\theta }{\,%
\left[ {\omega }^{2}\left( \omega ^{2}-2\,{\omega _{{0}}^{2}}+4\,{\theta }%
^{2}\right) +{\omega _{{0}}^{2}}\right] \,M}}\right) \,\tau ,
\]
an oscillating term,
\begin{eqnarray}
\fl J_{D,osc}(\tau ) &=&8\,{\frac{A\left( 3\,\theta \,{\omega }^{2}\,{\omega
_{{0}}}^{2}-4\,\theta \,{\ \omega _{{0}}^{4}}-2\,\theta \,{\omega }^{4}-4\,{%
\theta }^{3}\,{\omega }^{2}\right) {\bar{\Phi}}^{2}{\lambda _{{0}}^{2}}%
\gamma \sin \left( \omega \,\tau \right) }{{M}^{2}\omega \,\left( 4\,{\theta
}^{2}+{\omega }^{2}\right) \left( {\omega }^{4}+16\,{\ {\theta }^{2}\omega }%
^{2}+16\,{\omega _{{0}}^{4}}-8\,{\omega }^{2}{\omega _{{0}}^{2}}\right) }}+
\nonumber \\
&&2\,{\frac{A\left( 8\,\omega \,{\omega _{{0}}}^{4}+4\,{\theta }^{2}{\omega }%
^{3}+{\omega }^{5}-6\,{\omega }^{3}{\omega _{{0}}}^{2}\right) \,{\bar{\Phi}}%
^{2}{\lambda _{{0}}^{2}}\gamma \cos \left( \omega \,\tau \right) }{{M}%
^{2}\omega \,\left( {\omega }^{2}+4\,{\theta }^{2}\right) \left( {\omega }%
^{4}+16\,{\ \omega }^{2}{\theta }^{2}+16\,{\omega _{{0}}^{4}}-8\,{\omega }%
^{2}{\omega _{{0}}^{2}}\right) }}-  \nonumber \\
&&2\,{\frac{{A}^{2}\,\left[ {\omega \,\theta }^{2}\sin \left( 2\,\omega
\,\tau \right) +\left( \,{\omega _{{0}}^{2}}\,-{\omega }^{2}\right) \cos
\left( 2\,\omega \,\tau \right) \right] {\bar{\Phi}}^{2}{\lambda _{{0}}^{2}}%
\theta ^{2}\omega ^{2}}{M\,\,\left[ {\omega }^{2}\left( \omega ^{2}-2\,{%
\omega _{{0}}^{2}}+4\,{\theta }^{2}\right) +{\omega _{{0}}^{2}}\right] ^{2}}+%
}  \nonumber \\
&&{\frac{{A}^{2}\,\left[ {\omega }^{2}\left( {\omega }^{4}+-2\,{\omega _{{0}}%
}^{2}\right) +{\omega _{{0}}}^{4}\right] {\bar{\Phi}}^{2}{\lambda _{{0}}}%
^{2}\theta \,\omega \,\sin \left( 2\,\omega \,\tau \right) }{2\,M\,\,\left[ {%
\omega }^{2}\left( \omega ^{2}-2\,{\omega _{{0}}^{2}}+4\,{\theta }%
^{2}\right) +{\omega _{{0}}^{2}}\right] ^{2}}},
\end{eqnarray}
and a constant term,
\begin{eqnarray}
\fl J_{DTc} &=&{\frac{{\bar{\Phi}}^{2}{\ \lambda _{{0}}}}{\gamma }}-{\frac{{%
\bar{\Phi}}^{2}{\lambda _{{0}}^{2}}}{2\,M\,{\omega _{{0}}^{2}}}}+{\frac{A{%
\bar{\Phi}}^{2}{\lambda _{{0}}^{2}}\left( {\omega }^{2}-{\omega _{{0}}^{2}}%
\right) }{M\,\,\left[ {\omega }^{2}\left( \omega ^{2}-2\,{\omega _{{0}}^{2}}%
+4\,{\theta }^{2}\right) +{\omega _{{0}}^{2}}\right] }}+  \nonumber \\
&&{\frac{{A}^{2}\left[ \left[ 2\,{\omega }^{2}\left( {\omega _{{0}}^{2}}+2\,{%
\theta }^{2}\right) -{\omega _{{0}}^{4}}\right] {\omega _{{0}}^{2}}%
+8\,\omega ^{4}\,\left( {\theta }^{2}-{\omega _{{0}}}^{2}\right) \right] {%
\bar{\Phi}}^{2}{\ \lambda _{{0}}^{2}}}{2\,M\,\left[ {\omega }^{2}\left(
\omega ^{2}-2\,{\omega _{{0}}^{2}}+4\,{\theta }^{2}\right) +{\omega _{{0}%
}^{2}}\right] ^{2}}},
\end{eqnarray}
where $J_{DT}=J_{DT0}(\tau )+J_{D,osc}(\tau )+J_{DTc}$.

The injection of energy can be written in a similar way,
\begin{eqnarray*}
\fl J_{IT} &=&\int_{0}^{\tau }dt\,\int_{0}^{\infty }dt_{1}\,\delta
(t-t_{1})\,\int_{0}^{\infty }dt_{2}\,\delta (t-t_{2})\,v(t_{1})\,\eta
(t_{2}), \\
&=&\lim_{\epsilon \rightarrow 0}\int_{-\infty }^{\infty }\frac{dq_{1}}{2\pi }%
\,\int_{-\infty }^{\infty }\frac{dq_{2}}{2\pi }\,\frac{e^{(\mathrm{i}\,q_{1}+%
\mathrm{i}\,q_{2}+2\epsilon )\tau }-1}{(\mathrm{i}\,q_{1}+\mathrm{i}%
\,q_{2}+2\epsilon )}\,\tilde{v}(\mathrm{i}\,q_{1}+\epsilon )\,\tilde{\eta}(%
\mathrm{i}\,q_{2}+\epsilon ), \\
&=&\lim_{\epsilon \rightarrow 0}\int_{-\infty }^{\infty }\frac{dq_{1}}{2\pi }%
\,\int_{-\infty }^{\infty }\frac{dq_{2}}{2\pi }\,\frac{e^{(\mathrm{i}\,q_{1}+%
\mathrm{i}\,q_{2}+2\epsilon )\tau }-1}{(\mathrm{i}\,q_{1}+\mathrm{i}%
\,q_{2}+2\epsilon )}\,\left[ \frac{(\mathrm{i}\,q_{1}+\epsilon )}{M\,R(%
\mathrm{i}\,q_{1}+\epsilon )}\tilde{\eta}(\mathrm{i}\,q_{1}+\epsilon )\tilde{%
\eta}(\mathrm{i}\,q_{2}+\epsilon )\right] ,\,
\end{eqnarray*}
where after taking the thermal average gives,
\begin{eqnarray*}
\fl\left\langle J_{IT}\right\rangle &=&\left( \frac{\lambda _{0}\,\overline{%
\Phi ^{2}}}{M}\right) \lim_{\epsilon \rightarrow 0}\int_{-\infty }^{\infty }%
\frac{dq_{1}}{2\pi }\,\int_{-\infty }^{\infty }\frac{dq_{2}}{2\pi }\,\frac{%
e^{(\mathrm{i}\,q_{1}+\mathrm{i}\,q_{2}+2\epsilon )\tau }-1}{(\mathrm{i}%
\,q_{1}+\mathrm{i}\,q_{2}+2\epsilon )}\,\frac{(\mathrm{i}\,q_{1}+\epsilon )}{%
R(\mathrm{i}\,q_{1}+\epsilon )}\times \\
&&\left[ \frac{1}{\mathrm{i}\,q_{1}+\mathrm{i}\,q_{2}+2\,\epsilon }+\frac{A}{%
2}\left( \frac{1}{\mathrm{i}\,q_{1}+\mathrm{i}\,q_{2}+2\,\epsilon -\mathrm{i}%
\,\omega }+\frac{1}{\mathrm{i}\,q_{1}+\mathrm{i}\,q_{2}+2\,\epsilon +\mathrm{%
i}\,\omega }\right) \right. + \\
&+&\frac{\lambda _{0}}{2}\,\left\{ \frac{1}{\mathrm{i}\,q_{1}+\epsilon }+%
\frac{A}{2}\left( \frac{1}{\mathrm{i}\,q_{1}+\epsilon -\mathrm{i}\,\omega }+%
\frac{1}{\mathrm{i}\,q_{1}+\epsilon +\mathrm{i}\,\omega }\right) \right\}
\times \\
&\times &\left. \left\{ \frac{1}{\mathrm{i}\,q_{2}+2\,\epsilon }+\frac{A}{2}%
\left( \frac{1}{\mathrm{i}\,q_{2}+\epsilon -\mathrm{i}\,\omega }+\frac{1}{%
\mathrm{i}\,q_{2}+\epsilon +\mathrm{i}\,\omega }\right) \right\} \right] \\
&\equiv &J_{IT0}(\tau )+J_{IT,osc}(\tau )+J_{ITc}.
\end{eqnarray*}

An important part of the injection of energy flux has to be carefully
obtained since,
\begin{eqnarray*}
\left\langle J_{IT0}(\tau )\right\rangle _{1} &=&\frac{\lambda _{0}\,%
\overline{\Phi ^{2}}}{M}\lim_{\epsilon \rightarrow 0}\int_{-\infty }^{\infty
}\frac{dq_{1}}{2\pi }\,\int_{-\infty }^{\infty }\frac{dq_{2}}{2\pi }\,\frac{%
e^{(iq_{1}+iq_{2}+2\epsilon )\tau }-1}{(\mathrm{i}\,q_{1}+\mathrm{i}%
\,q_{2}+2\epsilon )^{2}}\,\frac{(\mathrm{i}\,q_{1}+\epsilon )}{R(\mathrm{i}%
\,q_{1}+\epsilon )} \\
&=&\frac{\lambda _{0}\,\overline{\Phi ^{2}}}{M}\,\tau \lim_{\epsilon
\rightarrow 0}\int_{-\infty }^{\infty }\frac{dq_{1}}{2\pi }\,\,\left( \frac{(%
\mathrm{i}\,q_{1}+\epsilon )}{(\mathrm{i}\,q_{1}+\epsilon -\kappa _{+})(%
\mathrm{i}\,q_{1}+\epsilon -\kappa _{-})}\right) , \\
&=&\frac{\lambda _{0}\,\overline{\Phi }^{2}}{\,M}\tau ,
\end{eqnarray*}
where the last term in the r.h.s. contains the integration over the the
upper arch, because the reduction lemma is not valid in this case, and using
the relations between raw moments of $\Phi $ as well.

Then finally, we write the contributions for the injection of energy as
\[
\fl J_{IT0}(\tau )=\left( \frac{\lambda _{0}\,{\bar{\Phi}}^{2}}{\,M}+{\frac{{%
A}^{2}{\omega }^{2}\,\theta \,{\lambda _{{0}}^{2}\,\bar{\Phi}}^{2}}{M\,\left[
{\omega }^{2}\left( \omega ^{2}-2\,{\omega _{{0}}^{2}}+4\,{\theta }%
^{2}\right) +{\omega _{{0}}^{4}}\right] }}\right) \,\tau ,
\]%
and
\begin{eqnarray}
\fl J_{IT,osc}(\tau ) &=&{\frac{\lambda _{{0}}{\bar{\Phi}}^{2}A\sin \left(
\omega \,\tau \right) }{M\omega }}+{\frac{A\left[ A\left( {\omega _{{0}}}%
^{2}-{\omega }^{2}\right) \,\cos \left( 2\,\omega \,\tau \right) +4\,{\omega
_{{0}}}^{2}\cos \left( \omega \,\tau \right) \right] {\lambda _{{0}}^{2}\,%
\bar{\Phi}}^{2}}{4\,M\,\left[ {\omega }^{2}\left( \omega ^{2}-2\,{\omega _{{0%
}}^{2}}+4\,{\theta }^{2}\right) +{\omega _{{0}}^{4}}\right] }}  \nonumber \\
&&+{\frac{A\left[ 2\omega \theta \left( A\sin \left( 2\,\omega \,\tau
\right) +4\sin \left( \omega \,\tau \right) \right) -4\,{\omega }^{2}\cos
\left( \omega \,\tau \right) \right] {\lambda _{{0}}^{2}\,\bar{\Phi}}^{2}}{%
4\,M\,\left[ {\omega }^{2}\left( \omega ^{2}-2\,{\omega _{{0}}^{2}}+4\,{%
\theta }^{2}\right) +{\omega _{{0}}^{4}}\right] },}
\end{eqnarray}%
and
\begin{eqnarray}
\fl J_{ITc} &=&{\frac{{\lambda _{{0}}^{2}\bar{\Phi}}^{2}}{M{\omega _{{0}}^{2}%
}}}+{\frac{A{\lambda _{{0}}^{2}\bar{\Phi}}^{2}\left( {\omega _{{0}}}^{2}-{%
\omega }^{2}\right) }{M\,\left[ {\omega }^{2}\left( \omega ^{2}-2\,{\omega _{%
{0}}^{2}}+4\,{\theta }^{2}\right) +{\omega _{{0}}^{4}}\right] }}+  \nonumber
\\
&&\,{\frac{\left[ {\omega }^{4}\left( {\omega }^{2}-12\,{\theta }^{2}\right)
+{\omega _{{0}}^{2}}\left( 3\,{\omega _{{0}}^{4}}-5\,{\omega _{{0}%
}^{2}\,\omega }^{2}+{\omega }^{4}-4\,{\theta }^{2}\right) \right] \,{A}^{2}{%
\lambda _{{0}}^{2}\,\bar{\Phi}}^{2}}{4\,M\,\left[ {\omega }^{2}\left( \omega
^{2}-2\,{\omega _{{0}}^{2}}+4\,{\theta }^{2}\right) +{\omega _{{0}}^{4}}%
\right] ^{2}}}.
\end{eqnarray}

For the total energy flux $J_{E}=J_{IT}+J_{DT}$ it can be easily seen that
the linear term on $\tau $ cancel out, while the constant term becomes
exactly the average energy of Eq.~(\ref{energiatotal}). The non-oscillating
part of the energy reads,
\begin{equation}
J_{E}^{no}=\,{\frac{{\bar{\Phi}}^{2}\lambda _{{0}}\left( 2\,M{\omega _{{0}}}%
^{2}+\lambda _{{0}}\gamma \right) }{2\gamma \,M{\omega _{{0}}^{2}}}}+{\frac{%
\left( {\omega }^{2}+{\omega _{{0}}}^{2}\right) {A}^{2}{\bar{\Phi}}^{2}{%
\lambda _{{0}}}^{2}}{4\,M\,\left[ {\omega }^{2}\left( \omega ^{2}-2\,{\omega
_{{0}}^{2}}+4\,{\theta }^{2}\right) +{\omega _{{0}}^{4}}\right] }},
\end{equation}
coinciding perfectly with Eq.~(\ref{energiatotal}). The oscillating term $%
J_{IT,osc}+J_{DT,osc}$ do not contribute if the time average is taken. Thus,
the average energy will in the form of an oscillation around the mean.

\begin{figure}[tbh]
\begin{center}
\includegraphics[width=0.60\columnwidth,angle=0]{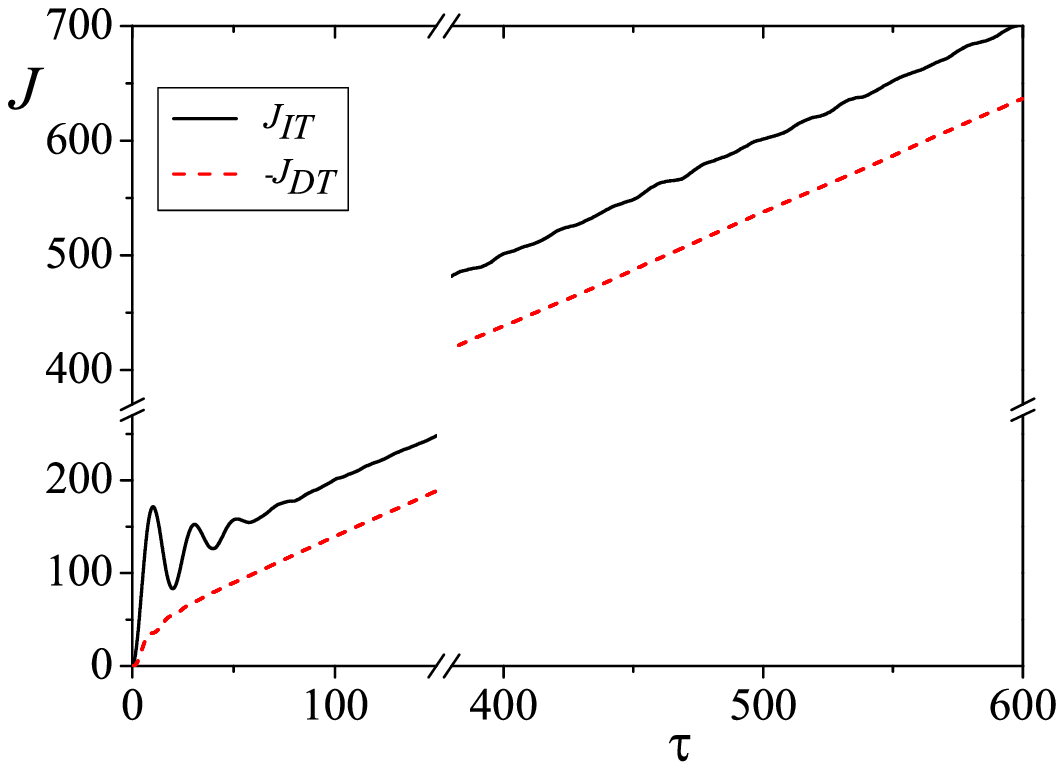} %
\includegraphics[width=0.60\columnwidth,angle=0]{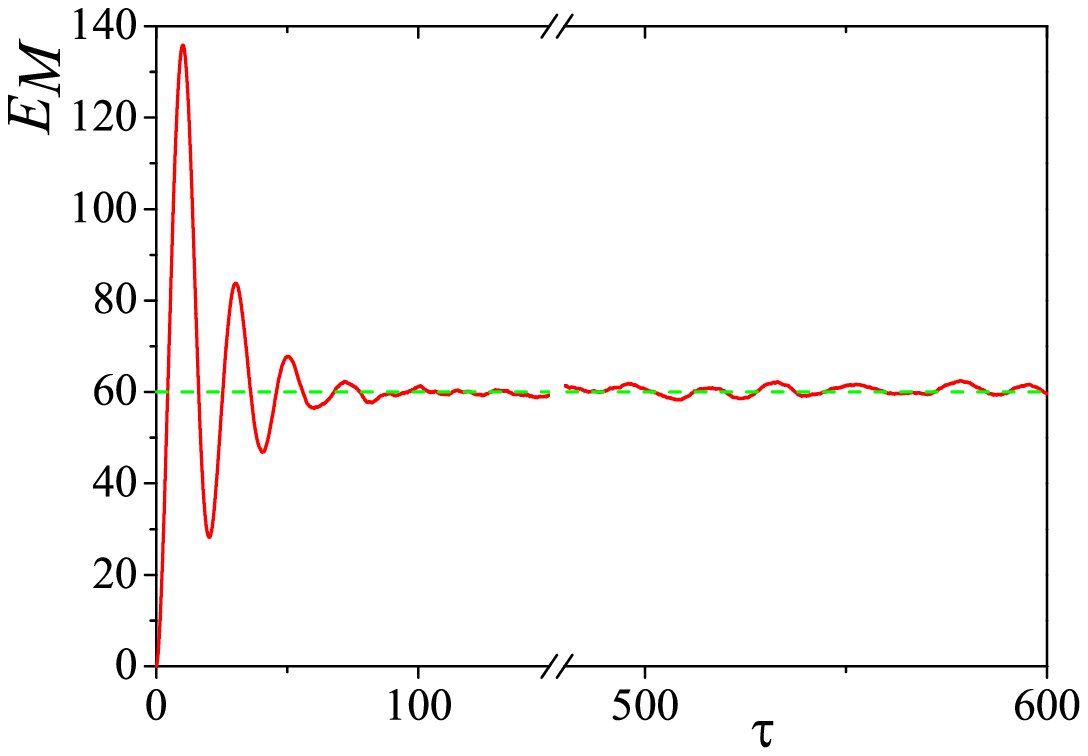}
\end{center}
\caption{Upper panel: Total injected and (symmetric) dissipated power, $%
J_{IT}\left( -J_{DT}\right) $ \textit{vs} time, $\protect\tau $, according
to the definitions in the legend. Lower panel: Total energy, $E_{M}$,
\textit{vs} time, $\protect\tau $. The dashed (green) line represents the
assymptotic limit given by Eq.~(\protect\ref{energiatotal}). In both cases
we have used the following values: $M=10$, $k_{0}=1$, $\protect\gamma =1$ $%
\bar{\Phi}=1$, $\protect\lambda _{0}=10$ and $A=0$. \emph{n.b.}: The
difference between the values of $J_{IT} $ and $-J_{DT} $ in the upper panel
is exactly equal to the total energy, $E_M$, that is shown in the lower
panel and equals the theoretical value given by Eq.~(\protect\ref%
{energiatotal}) as well.}
\label{Fig-Ja0}
\end{figure}

As expected, in the long-term the magnitudes of the injected and dissipated
energy fluxes are exactly the same signalling the emergence of equilibrium.
All of our calculations are compatible with the plots in Fig.~\ref{Fig-Ja0}
whereby we depict the evolution of the injected and dissipated power total
(mechanical) energy of an oscillator following our dynamical equations.

\begin{figure}[tbh]
\begin{center}
\includegraphics[width=0.65\columnwidth,angle=0]{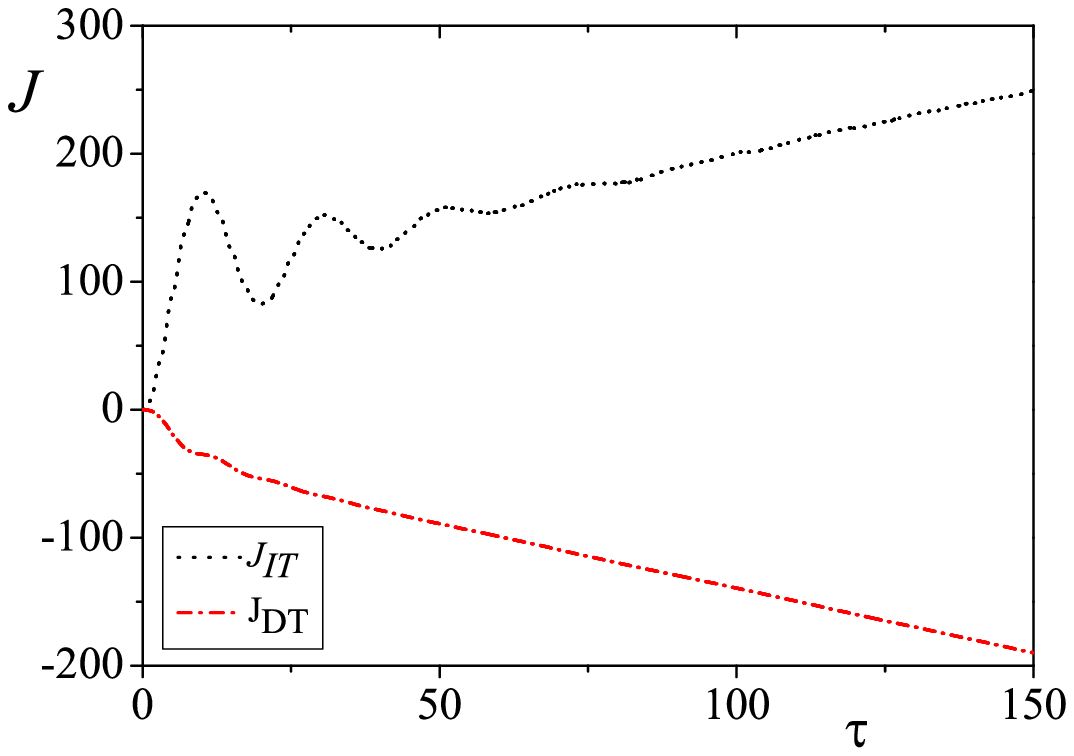} %
\includegraphics[width=0.65\columnwidth,angle=0]{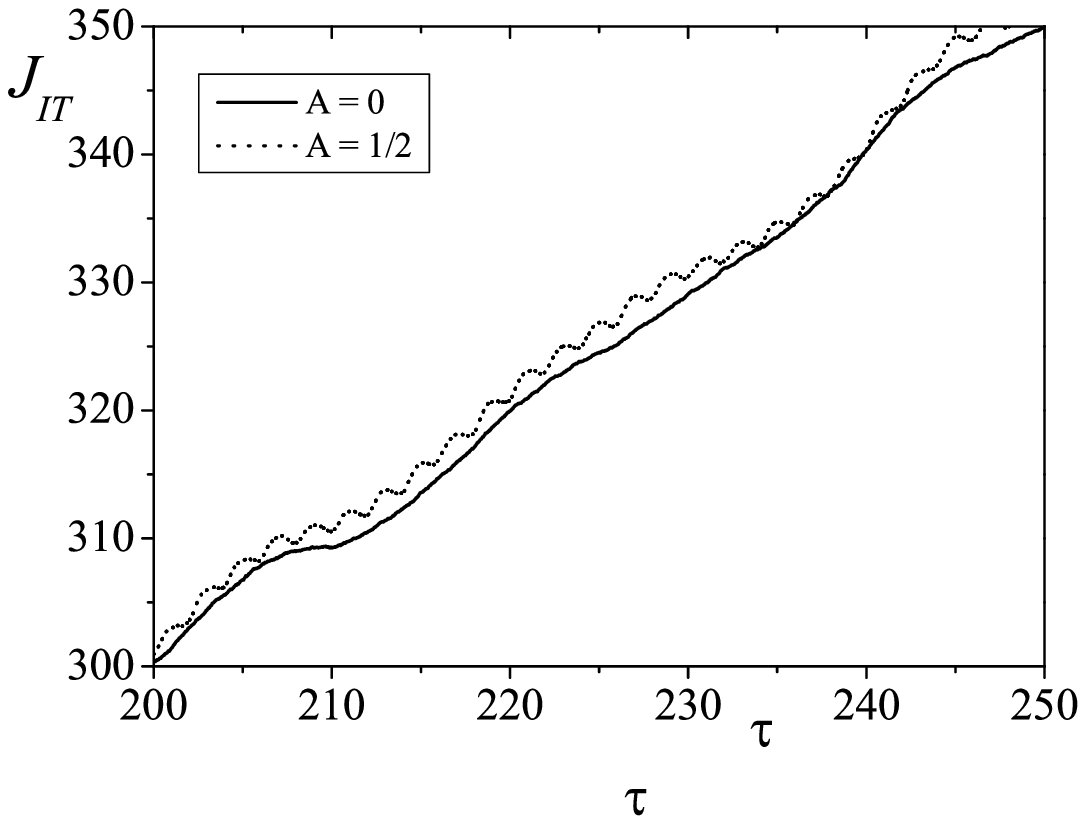}
\end{center}
\caption{Upper panel: Total injected, $J_{IT}$, and dissipated power, $%
J_{DT} $, \textit{vs} time, $\protect\tau $, according to the definitions in
the legend for the heterogeneous case $M=10$, $k_{0}=1$, $\protect\gamma =1$
$\bar{\Phi}=1$, $\protect\lambda _{0}=10$ and $A=1/2$. Lower panel:
Comparison of $J_{IT}$ for this heterogeneous process with a homogeneous
process ($A=0$) under the same remaining parameters. The impact of the
oscilations in the noise are clearly visible.}
\label{figKa05}
\end{figure}

\section{Concluding remarks}

In this work we have revisited the problem of the damped harmonic oscillator
subject to a heterogeneous Poisson process. Our approach, which is carried
out by averaging over the noise in the Fourier-Laplace space, allowed us to
obtain the long-term distributions of the position and distributions (joint
and marginal). Moreover, we have surveyed the interaction between the system
and the thermal bath by computing the rates of energy that are dissipated
from the system and injected into it. As expected, after a transient time,
both rates balance so that the system achieves a steady state. The
application of time averages over the position and velocity of the massive
partible has allowed obtaining the long-term distributions of the two
quantities, which is independent of the heteregeneous character of the
noise. This last feature will only have impact when averages over samples,
instead of averages of the time, are implemented. Notwithstanding, we having
been able to find the effect of the heterogeneity of the rate of events be
the emergence of resonance effects linking the ``natural'' frequency of
oscillation and and the frequency of the time-dependent part of the rate of
events. This impact is also visible when the total injected and dissipated
powers have been surveyed. Considerations regarding Jarzynski's equalities
as well as modifications on the inter-event rule of the noise are addressed
to future work.

\ack

WAMM would like to acknowledge Faperj and CNPq partial funding. SMDQ thanks
to E M F Curado and F D Nobre for their warm hospitality during short stays
at the Centro Brasileiro de Pesquisas F\'{\i}sicas sponsored by CNPq and
National Institute of Science and Technology for Complex Systems and the
financial support of the European Commission through the Marie Curie Actions
FP7-PEOPLE-2009-IEF (contract nr 250589) in the final part of the work. DOSP
would like to thank the Brazilian funding agency FAPESP for the financial
support.

\appendix

\section{Numerical calculations}

\label{appA}

In order to solve in a numerical way our main Eqs.~(\ref{01})~and~(\ref{02})
we have considered a trapezoidal approximation which is reminiscent of the
Stratanovich approach to noise phenomena,%
\[
x\left( t_{2}\right) -x\left( t_{1}\right) =\int_{t_{1}}^{t_{2}}v\left(
t\right) \,dt\approx \left( t_{2}-t_{1}\right) \frac{v\left( t_{2}\right)
+v\left( t_{1}\right) }{2},
\]%
and the same for the deterministic part of the velocity,%
\begin{eqnarray*}
\Delta v^{\det }\left( t_{2},t_{1}\right) &\equiv
&M^{-1}\int_{t_{1}}^{t_{2}} \left[ -k_{0}x(t)-\gamma v(t)\right] dt \\
&\approx &\left( t_{1}-t_{2}\right) \frac{\gamma \left[ v\left( t_{2}\right)
+v\left( t_{1}\right) \right] +k_{0}\left[ x\left( t_{2}\right) +x\left(
t_{1}\right) \right] }{2\,M},
\end{eqnarray*}%
for small enough $dt$.

In respect of the stochastic part~\cite{2007_PRE_76_011109}, its calculation
can be at least made threefold. The first one concerns the inter-event time
which is given by Eq.~(\ref{pdftempo}). Accordingly, starting from $t_{0}$
we would randomly select a certain time interval, $\delta t$, following PDF
in Eq.~(\ref{pdftempo}) and we would let deterministic equations evolve up
to $t_{0}+\delta t$ when we would add the value of the kick, $M^{-1}\Phi
\left( t_{0}+\delta t\right) ,$ to $\Delta v^{\det }$ with $\Phi $ chosen
from the specific distribution $P\left( \Phi \right) $. Despite its
accurateness this procedure is not the most hard-headed when it comes to
simulating heterogeneous Poisson processes, since we are obliged to
constantly update the distribution in a rather grinding way.

The second method is a shrewd procedure of carrying out the numerical
simulation without having to heed to a very tiny value of the mesh $\delta t$
or also to problems with a very high rate of occurrence of events $\lambda
\left( t\right) $. In this case we can determine the (expected) number of
events that take place within $\delta t$,%
\[
n\left( t\right) =\int_{t}^{t+\delta t}\lambda \left( t^{\prime }\right)
\,dt^{\prime },
\]%
and consider that the overall effect of the noise in that time interval is
equivalent to the occurrence of a single kick the intensity of which is
given by the convolution of $n\left( t\right) $ distributions $P\left( \Phi
\right) $. Bearing in mind that the events are uncorrelated the resulting
distribution is given by,%
\[
P_{n}\left( \Phi \right) =\mathcal{F}_{\Phi }^{-1}\left[ \mathcal{F}_{\phi }%
\left[ P\left( \Phi \right) \right] ^{n}\right] .
\]%
However, it must be stressed that this procedure is half-averaged since it
already assumes the mean number of events in its implementation, thus
leaving all the randomness to the resulting amplitude of the added noise.
Although we have not tested the following assertation, we believe that its
application reduces the number of samples needed to obtain the same
dispersion in the sample set.

The third way corresponds to our main option, particularly for the figures
in the text. Specifically, we were intentionally careless about optimising
computational time, as we have preferred a very conservative approach and a
very tight grid. Since we did not opt for depicting examples with very high
event rates, we went ahead by picking a random number uniformly distributed
between 0 and 1 and compared it with the probability of having an event
according with the Poisson distribution with parameter $\int_{t}^{t+\delta
t}\lambda \left( t^{\prime }\right) \,dt^{\prime }$, which is similar to
Eq.~(\ref{lambdaintegration1}). If the random number is the smaller of both
numbers, then a kick takes place and therefore we need to select the noise
intensity as described for the first approach. In all the cases we have
shown $\delta t=10^{-4}$. The distributions were obtained from a total of $%
5\times 10^{8}$ records (10$^{3}$ samples) made at intervals of $10^{-3}$
time units. To assure equilibrium we have set apart the first 10$^{5}$ (100
time units) logs of each sample.

\end{document}